\documentclass[aps,prl,twocolumn,superscriptaddress,showpacs,floatfix]{revtex4-1}
\usepackage{graphicx,amssymb}
\usepackage[fleqn]{amsmath}
\usepackage{amsfonts}
\usepackage{dcolumn}
\usepackage{bm}
\usepackage{braket}
\usepackage{multirow} 
\usepackage{MnSymbol}

\graphicspath{{images/}}

\newcommand{\hatvec}[1]
{\hat{\vec{#1}}}

\renewcommand{\vec}[1]{\mbox{\boldmath $#1$}}

\begin{document}

\date{\today}

\title{Nuclear rotation in the continuum}

\author{K. Fossez}
\affiliation{NSCL/FRIB Laboratory,
Michigan State University, East Lansing, Michigan  48824, USA}

\author{W. Nazarewicz}
\affiliation{Department of Physics and Astronomy and NSCL/FRIB Laboratory,
Michigan State University, East Lansing, Michigan  48824, USA}
\affiliation{Institute of Theoretical Physics, Faculty of Physics, 
University of Warsaw, Warsaw, Poland}

\author{Y. Jaganathen}
\affiliation{NSCL/FRIB Laboratory,
Michigan State University, East Lansing, Michigan  48824, USA}

\author{N. Michel}
\affiliation{Grand Acc\'el\'erateur National d'Ions Lourds (GANIL), CEA/DSM - CNRS/IN2P3,
BP 55027, F-14076 Caen Cedex, France}

\author{M. P{\l}oszajczak}
\affiliation{Grand Acc\'el\'erateur National d'Ions Lourds (GANIL), CEA/DSM - CNRS/IN2P3,
BP 55027, F-14076 Caen Cedex, France}

\begin{abstract}
	\begin{description}
		\item[Background]
			Atomic nuclei often exhibit collective rotational-like  behavior in highly excited states, well above the particle emission threshold.
			What determines the existence of collective motion in the  continuum region, is not fully understood.
		\item[Purpose]
			In this work, by studying the collective rotation of the positive-parity deformed configurations of the one-neutron halo nucleus  $^{11}$Be, we assess different mechanisms that stabilize collective behavior beyond the limits of particle stability.
		\item[Method]
			To solve a particle-plus-core problem, we employ a non-adiabatic coupled-channel formalism and the Berggren single-particle ensemble, which explicitly contains bound states, narrow resonances, and the scattering continuum.
			We study the valence-neutron density in the intrinsic  rotor frame to assess the validity  of the adiabatic approach as the excitation energy increases.
		\item[Results]
			We demonstrate that collective rotation of the ground band of $^{11}$Be is stabilized by (i) the fact that the $\ell=0$ one-neutron decay channel is closed, and (ii) the angular momentum alignment, which increases the parentage of high-$\ell$ components at high spins; both effects act in concert to decrease decay widths of ground-state band members.
			This is not the case for higher-lying states of $^{11}$Be, where the $\ell=0$ neutron-decay channel is open and often dominates.
		\item[Conclusion]
			We demonstrate that long-lived collective states can exist at high excitation energy in weakly bound neutron drip-line nuclei such as $^{11}$Be. 
	\end{description}
\end{abstract}


\maketitle

%
%
\textit{Introduction.} --  Studies of exotic nuclei far from the valley of beta-stability reveal novel features, such as the formation of halo structures \cite{jensen04_233,tanihata13_549}, near-threshold clustering effects \cite{oertzen06_1017,freer07_1018,okolowicz13_241,okolowicz12_998}, and presence of new types of correlations \cite{Matsuo10}.
In all these cases, the atomic nucleus exhibits properties characteristic of an open quantum system, whose properties are dramatically altered by the coupling to scattering and reaction channels \cite{michel10_4}.

While the impact of reaction channels  on nuclear structure has been recognized \cite{michel09_2}, the fact that some highly excited nuclear excitations can be interpreted in terms of nuclear clusters and nuclear molecules \cite{oertzen06_1017,freer07_1018} that experience  collective motions such as rotations and vibrations, is quite astonishing.
The success of the collective model of atomic molecules relies on the validity of the  adiabatic Born-Oppenheimer approximation \cite{born27} that relies on the vast difference between time scales for single-electron and ionic motions.
However, when it comes to atomic nuclei, the time separation between single-nucleon and collective  motion is small, and the adiabatic approximation is expected to be badly violated \cite{naz93}.
How come, therefore,  that a highly excited state, undergoing rapid particle emission, can  be viewed in simple geometric or algebraic terms involving rotating  or vibrating fields, or potentials, common for all nucleons?
An excellent recent example is offered by the spectrum of $^{12}$C discussed in the geometric/algebraic language of  three $\alpha$-particle arrangements \cite{Lambarri14}.

Whether a broad feature observed in scattering experiments can be understood in terms of a nucleus, or a nuclear state, is a subject to ongoing debate.
Consider the single-particle time scale.
The  average  time it  takes a nucleon   to go across a light nucleus (${ A \approx 10 }$) and come back can be roughly estimated at $T_{\rm s.p.} \approx 1.3{\cdot}10^{-22}$\,sec \cite{goldanskii66_1159}, and corresponds to the time scale needed to create the nuclear mean field.
If one relates the half-life of the unbound nuclear state to its decay width $\Gamma$ through $T_{1/2}=\ln(2)\hbar/\Gamma$, one is tempted to conclude that broad scattering features with $T_{1/2}< T_{\rm s.p.}$ (or $\Gamma > 3.5$\,MeV for $A\approx 10$) can hardly be interpreted in terms of nuclear states \cite{thoennessen04_1165}. 
Of course, there is no sharp borderline that separates {\it bona fide} nuclear states from broad features seen in scattering experiments, and this often results in interpretational difficulties  \cite{garrido13_1171}.
To cloud the issue even more,  a quantitative experimental characterization of broad resonances, embedded in  a large non-resonant background,  is  not always possible.
For instance, the extraction of experimental widths is usually model-based and relies on approximations \cite{Fynbo09,Rii15}.
As discussed in Ref.~\cite{garrido13_1171}, many theoretical descriptions of collective states in the continuum \cite{pastore14_1183,caprio13_1158} employ bound-state techniques using localized wave functions; those have limited applicability when it comes to unbound states and broad  resonances in particular. 

The purpose of this work is to shed light on the notion  of nuclear collective motion in the continuum.
Ideally, to address this question, one should use a  microscopic many-body approach that allows the cluster degrees of freedom to naturally emerge from the $A$-body problem in the open-system formulation \cite{hagen06_14,quaglioni12_738,papadimitriou13_441}.
The purpose of this work, however, is not to solve the  many-body problem in its full glory with the inclusion of continuum space, but rather to gain insight into the notion of collective resonances  by inspecting a simple case.
To this end, we use a schematic particle-plus-core coupled-channel approach, based on the Berggren ensemble \cite{berggren68_32}.
This model   explicitly contains bound states, resonances and complex-energy scattering states, i.e., all ingredients needed to describe the coupling to the one-nucleon continuum, and it also allows for a simple explanation in terms of familiar geometrical terms.
In a similar context, this approach has been applied successfully to the description of the resonant spectrum of the dipolar anion \cite{fossez15_1028}, which can be considered as an extreme one-electron halo.
It has been shown that below the dissociation threshold, the motion of the valence electron  is strongly coupled to the collective rotation of the molecule, forming rotational states.
Above the ionization threshold, however,  a rapid decoupling of the electron's motion from the  rotational motion of the  molecule takes place, thus leading to a disappearance of a collective rotational band.
This observation begs the question whether such a demise of collective rotation could also be expected in nuclear halo systems.
Here, we investigate qualitatively the existence of nuclear rotational states in the continuum of the deformed one-neutron halo nucleus  $^{11}$Be.

\textit{Model and parameters} -- To solve a problem of a particle coupled to a collective core, we apply the non-adiabatic coupled-channel method of Refs.~\cite{kruppa00_581,barmore00_582,kruppa04_472,esbensen00_386} originally developed in the context of deformed proton emitters.
We assume that the positive-parity ground-state band of $^{11}$Be can be viewed as a weakly-bound/unbound neutron coupled to a deformed core of $^{10}$Be \cite{nunes96_1127,Descouvemont1997261}. The exact form of the deformed  pseudo-potential representing the neutron-core interaction is not essential for the purpose of our qualitative discussion, as long as the one-neutron threshold is reproduced.
Here we  approximate it by a deformed Woods-Saxon (WS) potential with a spherical spin-orbit term \cite{barmore00_582}.
The total angular momentum of the system is ${ \hatvec{J} = \hatvec{j} + \hatvec{j}_r }$, where
${ \hatvec{j} = \hatvec{\ell} + \hatvec{s} }$ is the anglar momentum of the valence nucleon and 
${ \hatvec{j}_r }$ is that of the rotor.

In the coupled-channel (CC) formalism, eigenstates of the decaying nucleus ${ \ket{ \Psi^{ J^\pi } } }$ are expanded in the  basis of channel wave functions labeled by channel quantum numbers ${ c = ( \ell j j_r ) }$.
Each channel state is given by the cluster radial wave function $u_c (r) / r$ representing the relative radial motion of the particle and the core, and the orbital-spin part  ${ \ket{ j ( \ell , s ) j_r ; J  M_J } }$.

The CC equations for $u_c (r)$ are:
\begin{eqnarray}\label{eq_CC_eqs}
  \left[-\frac{\hbar^2}{2\mu} \frac{d^2}{dr^2} \right.
& + & \left. \frac{\hbar^2 \ell(\ell+1)}{2\mu r^2} + 
  V_{c,c}^J(r) - Q(J^\pi,j_r) \right]  u_c^{ J^\pi } (r)  \nonumber\\
 & + & \sum_{c'\neq c} V_{c,c'}^J(r) \;
u_{ c' }^{ J^\pi } (r)=0,
\end{eqnarray}
where  $V_{c,c}^J$ is the diagonal part of the neutron-core potential, $Q(J^\pi,j_r)=E^{ J^\pi }- E_d^{j_r^\pi}$ is the energy of the  particle decaying from the state $E^{J^\pi}$ of the parent nucleus to the state $E_d^{j_r^\pi}$ of the daughter system, and $V^{J}_{c,c'}(r)$ are the off-diagonal coupling terms.  

The CC equations \eqref{eq_CC_eqs} can be solved using the Berggren expansion method (BEM) \cite{fossez13_552,fossez15_1028}. 
This approach has been benchmarked in Ref.~\cite{fossez13_552} by using the traditional technique of direct integration of CC equations (DIM), which is fairly accurate for well localized states, but less accurate for unbound Gamow states or when a large number of channels is considered. 
Indeed, the DIM and BEM results are very close for the bound ground state and low-lying narrow resonances, but for broader resonant states the BEM predicts smaller energies and widths; hence, it is favored by the generalized variational principle.
Moreover, the DIM requires a starting point for the integration that is close to the final eigenenergy, otherwise it can sometimes exhibit a divergent behavior, and the search for a good starting point has to be carried out for all individual eigenstates of interest, while in the BEM the whole spectrum is obtained in one diagonalization.
Finally, BEM calculations  are much faster, especially for a large number of channels.
For instance, for the ${ {7/2}^{+} }$ yrast state of ${ {}^{11}\text{Be} }$ with ${ {\ell}_{ \text{max} } }=6$ (22 channels),  the time required by the DIM to solve the CC equations is over three orders of magnitude longer than in the BEM.

In the present case, a Berggren basis is built for each partial wave $(\ell, j)$, from the discrete solutions of the diagonal part of Eq.~\eqref{eq_CC_eqs}  with outgoing boundary conditions.
The off-diagonal matrix elements $V^{J}_{c,c'}$  are calculated using the exterior complex-scaling technique \cite{dykhne61_1041,gyarmati71_38}.

The rotational structure of nuclear states can be interpreted in terms of the internal density of the valence particle in the core reference frame.
Here we assume that the core is associated with the rigid rotor axially deformed around the  $z$-axis.
When expressed in the deformed reference frame, the eigenstates  ${ \ket{ \Psi^{ J^\pi } } }$ can be expanded in the basis $\ket{ \Psi^{ J^\pi }_K }$, where ${ K }$ is the projection of the total angular momentum ${ J }$ on the symmetry axis in the core frame.
The density ${ { \rho }_{ J K } ( r , \theta ) }$ is  calculated as an average over all orientations of  the density operator \cite{fossez15_1028}.
The total density is then obtained by summing up of all $K$-components.
If only one ${ K }$-component is nonzero, ${ K }$ becomes a good quantum number and the strong coupling (adiabatic) limit is strictly obeyed \cite{kruppa00_581,barmore00_582,esbensen00_386,davids04_562}; the resulting density is referred to as the \textit{intrinsic density}.
%

Coupled-channel equations were solved  up to a maximal radius of $R_\text{max}= 30$\,fm and the rotation radius for the exterior complex-scaling $R_\text{rot} = 15$\,fm.
Complex-energy scattering states entering in the Berggren basis are selected along a contour in the complex momentum plane defined by three segments connecting the points (0,0), (0.2,$-$0.2), (0.5,0), and (2,0); all in fm$^{-1}$.
The contour has  been discretized with 60 points using a Gauss-Legendre quadrature.
Since the rotational bands in question have positive parity, we took  partial waves with $\ell=0, 2,4, 6$ in the Berggren ensemble, and the maximum angular momentum of the core that was large enough to guarantee that the number of included states in the ground-state band of the daughter nucleus would not impact the calculated widths \cite{barmore00_582,fossez15_1028}.
In order to partly satisfy the Pauli principle between core and valence particles, the core neutron $0s_{1/2}$ shell   has been excluded in the construction of the CC basis \cite{saito69}.
We checked that with this choice of parameters, the solutions of CC equations are perfectly stable.

For the energies of the core nucleus  $E_d^{j_r^\pi}$, we took the known experimental $j_r^\pi = 0^+, 2^+$, and 4$^+$  members of the ground-state band of $^{10}$Be \cite{ensdf}.
The higher-lying band members are approximated by means of the rigid rotor expression with the moment of inertia  corresponding to the 4$^+$ level.
In the following discussion, we ignore the particle width of the core states with $j_r\ge 4$.
(Experimentally the $2^+_1$ state in $^{10}$Be is particle-bound, and the $4^+_1$ level has a fairly small width of 121\,keV.)

The  parameters of the deformed pseudo-potential have been fitted to the $1/2^+$ and 5/2$^+$ members of the  yrast band of $^{11}$Be \cite{ensdf}.
We have checked that  these states collapse to the same band-head energy in the adiabatic limit (${ I \to \infty }$), i.e.,  they are members of the same rotational band.
It is worth noting that the location of higher-lying band members is very uncertain, although  candidates for the $3/2_1^+$, 7/2$_1^+$, and 9/2$_1^+$ states have been suggested \cite{voer98,Bohlen08}.
The optimized parameters of the WS pseudo-potential are:  radius $R_0=2.08$\,fm; diffuseness $a=1.1$\,fm; strength $V_0=-59.36$\,MeV; spin-orbit strength $V_{\rm so}=15.09$\,MeV; and quadrupole deformation $\beta_2=0.53$. We have checked that the general conclusions of our study are independent on the precise values of these parameters, as long as the one-neutron threshold is reproduced. For instance, similar  results were obtained using ${ a = 0.77 }$\,fm, ${ R_0 = 2.55 }$\,fm, and ${ \beta_2 = 0.52 }$.

\begin{figure}[htb]	\includegraphics[width=0.90\linewidth]{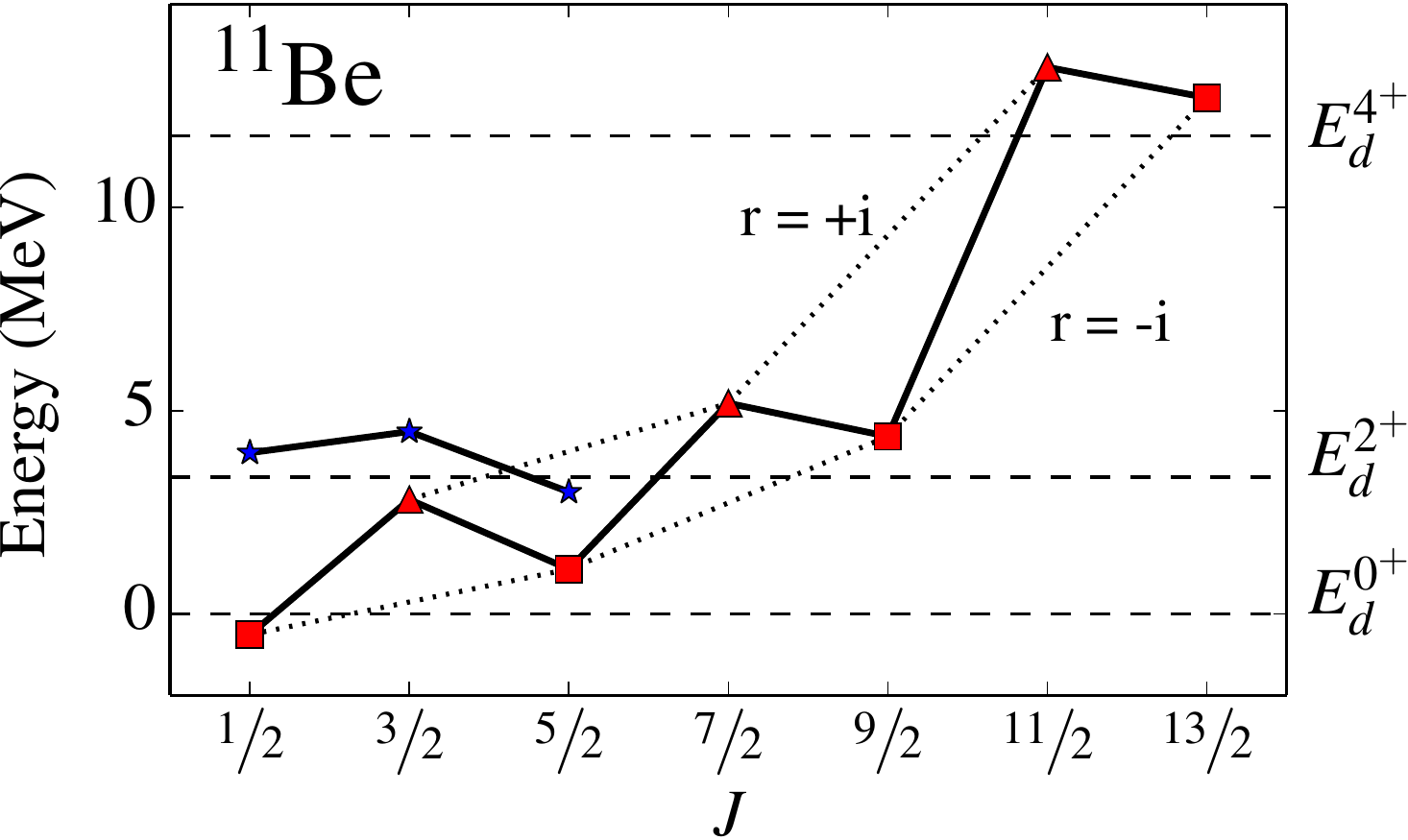}
	\caption{(Color online)  Calculated yrast bands of  $^{11}$Be with $J\le 13/2$
	with signature $r=-i$ (squares) and $r=+i$ (triangles)
	compared to the experimental ground-state band of $^{10}$Be core (horizontal dashed lines). Some excited (yrare) states of $^{11}$Be are marked by stars.}
	\label{fig_1}
\end{figure}
\textit{Results} -- The calculated lowest states of $^{11}$Be  are shown in Fig.~\ref{fig_1}. The large splitting between the favorite-signature band ($r=\exp(-i\pi J)=-i$) and unfavored band ($r=+i$) is consistent with the results of a microscopic multicluster model \cite{Descouvemont1997261,descouvemont02_1129} and large-scale shell model \cite{caprio13_1158,maris15_1161}.
It is interesting to note that $Q(J,j_r)<0$ for $|J-j_r|=1/2$ for the  yrast states in $^{11}$Be.
For instance, the $3/2_1^+$ and $5/2^+_1$ levels of $^{11}$Be are predicted to lie {\it below} the yrast $2^+$ state of $^{10}$Be.
This means that the $\ell=0$ neutron emission channel is blocked for both $r=-i$ and $r=+i$ bands.

The rotational structure of the ground-state band of $^{11}$Be is revealed by looking at  the  weights of individual ${ K }$-components of valence neutron density.
It turns out that the ${ K = 1/2 }$ component is dominant in most cases; in particular for the favored band.
For the 7/2$^+$, 11/2$^+$, and  15/2$^+$ states, the ${ K = 5/2 }$ and ${ 3/2 }$ components dominate.
In most cases, an appreciable degree of $K$-mixing is predicted.
This suggests that a ``$K=1/2$" label often attached to this band should be taken with a grain of salt.

\begin{figure}[htb]
\includegraphics[width=0.80\linewidth]{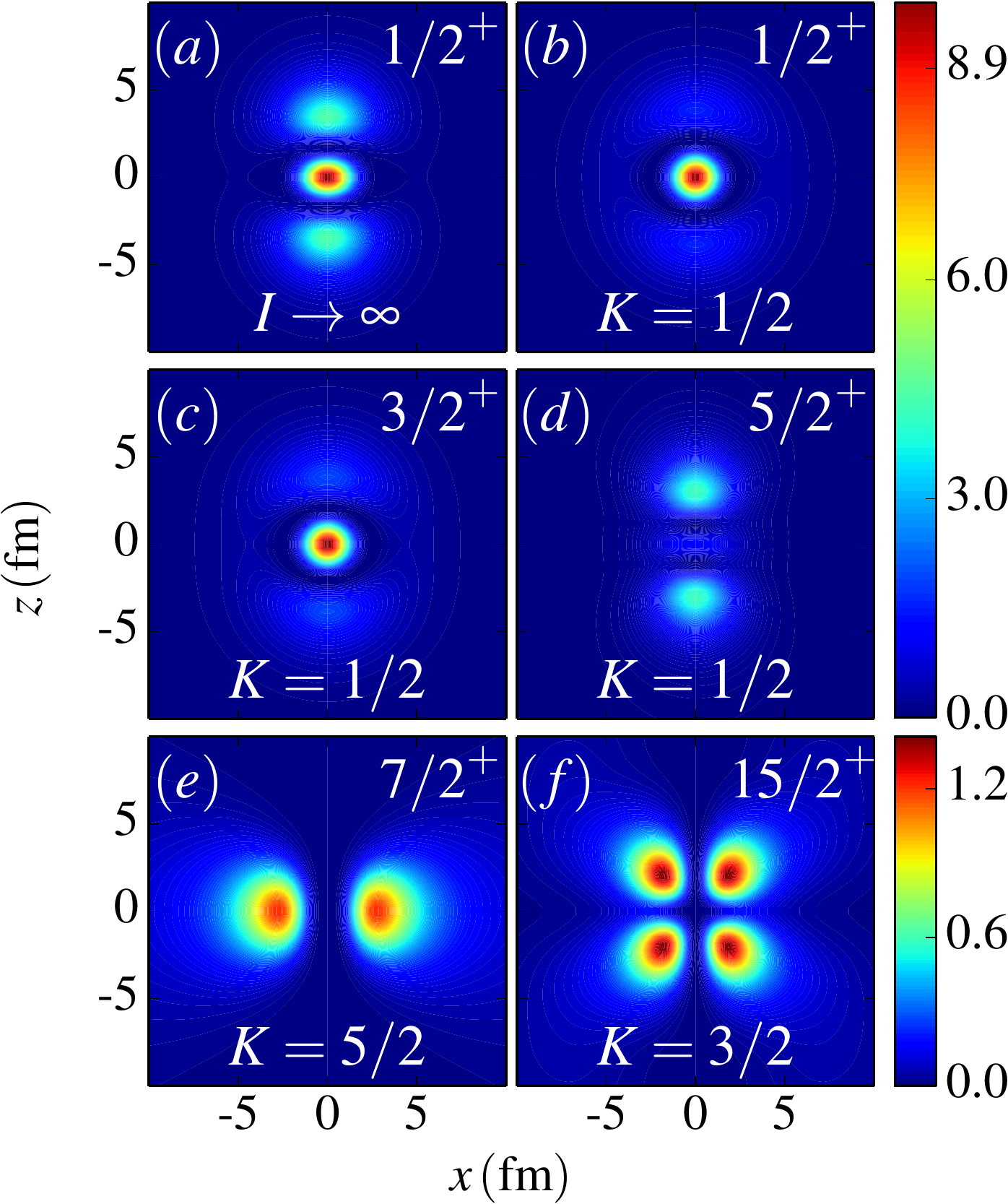}
\caption{(Color online)  Partial densities ${ \rho_{ J K } ( \vec{r}) }$ (in $10^{-2}$\,fm$^{-3}$) with ${ K = 1/2 }$ for the three first ground-state band members of $^{11}$Be in panels (b), (c) and (d), and for selected states of the unfavored $r=i$ band (in $10^{-3}$\,fm$^{-3}$) in panels (e) $J=7/2, K=5/2$ and (f) $J=15/2, K=3/2$. The adiabatic limit ($I\rightarrow \infty$) is shown in panel (a).}
	\label{fig_2}
\end{figure}
%
To get further insight into the structure of the  yrast line of $^{11}$Be, the density ${ { \rho }_{ J K } (\vec{r}) }$ of the valence neutron in the core frame is plotted in Fig.~\ref{fig_2} for selected states and ${ K }$-components.
In the adiabatic limit, we checked that all band members have indeed the same ${ K = 1/2 }$ intrinsic density shown in Fig.~\ref{fig_2}(a), whereas all ${ K > 1/2 }$ components vanish.
For the favored band, for which the $K=1/2$  component dominates,  the densities 
${ { \rho }_{ J K=1/2 } }$
for $J=9/2, 13/2$, and $17/2$ are similar to that for the ${ {5/2}^+ }$ state shown in Fig.~\ref{fig_2}(d).
This is not the case for the ${ {7/2}^+ }$, $11/2^+$, and $15/2^+$ band members, which show an appreciable $K$-mixing. Indeed, as illustrated
in Figs.~\ref{fig_2}(e) and (f), the dominant  neutron distributions for
the $7/2^+$ and $15/2^+$ states are very different.
%
Figure~\ref{fig_2} also shows that the  calculated valence neutron  states in ${^{11}}$Be are all fairly well localized within the range of the neutron-core potential. We checked that this  holds  for the yrare states as well.

\begin{figure}[htb]
\includegraphics[width=0.90\linewidth]{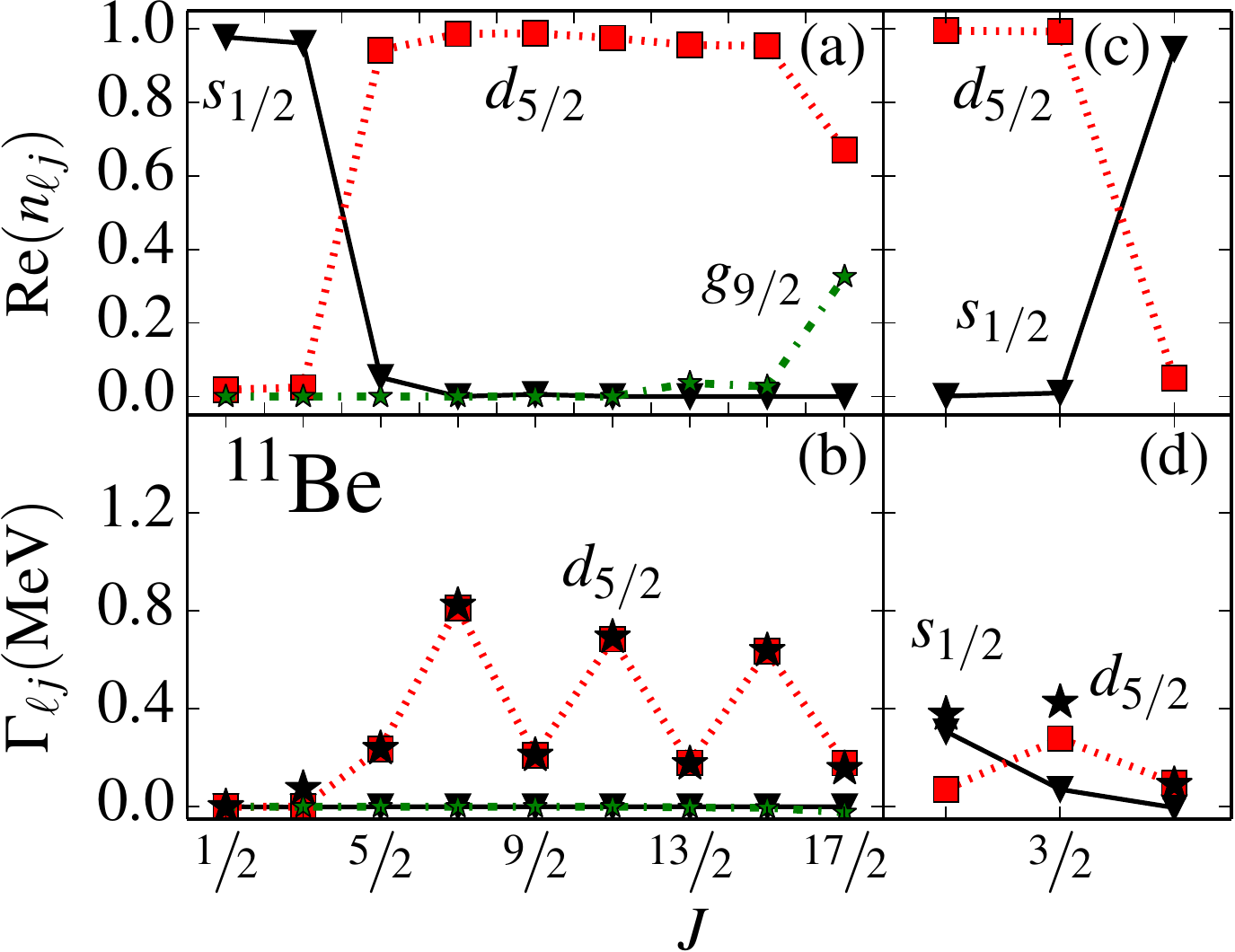}
	\caption{(Color online)  Contribution from different  partial waves ${ (\ell  j ) }$ to the  norms (top) and widths $\Gamma_{\ell j}=\sum_{j_r}\Gamma_{\ell j j_r}$ (bottom) of different states of the ground-state band  of $^{11}$Be (left) and yrare states (right). The total widths are marked by stars.}
	\label{fig_3}
\end{figure}

Figure~\ref{fig_3}(a) shows the (real part) of the norms $n_{\ell j}=\sum_{j_r}n_{\ell j j_r}$,  from various partial waves $(\ell j)$ to the yrast band of $^{11}$Be.
It is seen that the alignment pattern of the valence neutron is governed by a transition from the $s_{1/2}$ wave, which dominates at low spins, to $d_{5/2}$, which governs the rotation at higher angular momenta.
In the high-spin region, $J\ge 7/2$, the yrast line of $^{11}$Be can be associated with the weak coupling of neutron's angular momentum, $j=5/2$, to the angular momentum $j_r$ of the core resulting in the full rotational alignment of $\vec{j}$ with $\vec{j}_r$.
Indeed, as seen in Fig.~\ref{fig_1}, the computed energies $ E^{J}$ of the $r=-i$ band appear close to the energies of $^{10}$Be with $j_r=J-5/2$.
The higher partial waves, such as $g_{9/2}$, appear for ${ J > 15/2 }$.
The structure of yrare states shown in Fig.~~\ref{fig_3}(c) is dominated by $d_{5/2}$ at low spins, and by $s_{1/2}$ at higher angular momenta.

To estimate one-neutron decay rates, we compute  decay widths.
An accurate way to compute the 
total decay width is to use  the so-called current expression \cite{humblet61_174,barmore00_582,kruppa04_472}, that gives $\Gamma=\sum_{c} \Gamma_{c} (r)$ in terms of partial widths from the channel wave functions.
We checked that in every case considered the value of $\Gamma$ obtained from the current expression agrees with the eigenvalue estimate  $-2{\rm Im}(E^{J^\pi})$ and that the values of $\Gamma_c (r)$ are stable at  $r=R_{\rm max}$.

The calculated one-neutron  widths corresponding to different partial waves are shown in panels (b) and (d) of Fig.~\ref{fig_3}.
As discussed earlier, the $s$-wave neutron decay  is blocked in the ground band of $^{11}$Be.
Consequently, the neutron widths of yrast states are primarily governed by $\ell=2$ waves.
Moreover, because of the weak coupling of the valence neutron to the core states, the $Q$-values for the neutron decay of the favorite band is small in the $d_{5/2}$ channel.
As a result, the states in the favorite yrast band of $^{11}$Be are predicted to have  small neutron widths of the order of 200\,keV.
In the case of the unfavored band, the neutron widths are larger, around $\Gamma=0.7$\,MeV.
This  result demonstrates that angular momentum alignment can stabilize collective behavior in highly excited yrast states of a neutron drip-line system.
The underlying mechanism is similar to that discussed in Ref.~\cite{nazarewicz01_1180} in the context of the cranking description of rotational properties of neutron drip-line nuclei.
Due to  the Coriolis force,  high-$\ell$ orbits responsible for angular momentum alignment become  occupied at high spins at the expense of  low-$\ell$ states.
The latter  govern halo properties and particle decay; the former are well localized within the nuclear volume because of their large centrifugar barrier. 

\begin{figure}[htb]
		\centering
		\includegraphics[width=0.90\linewidth]{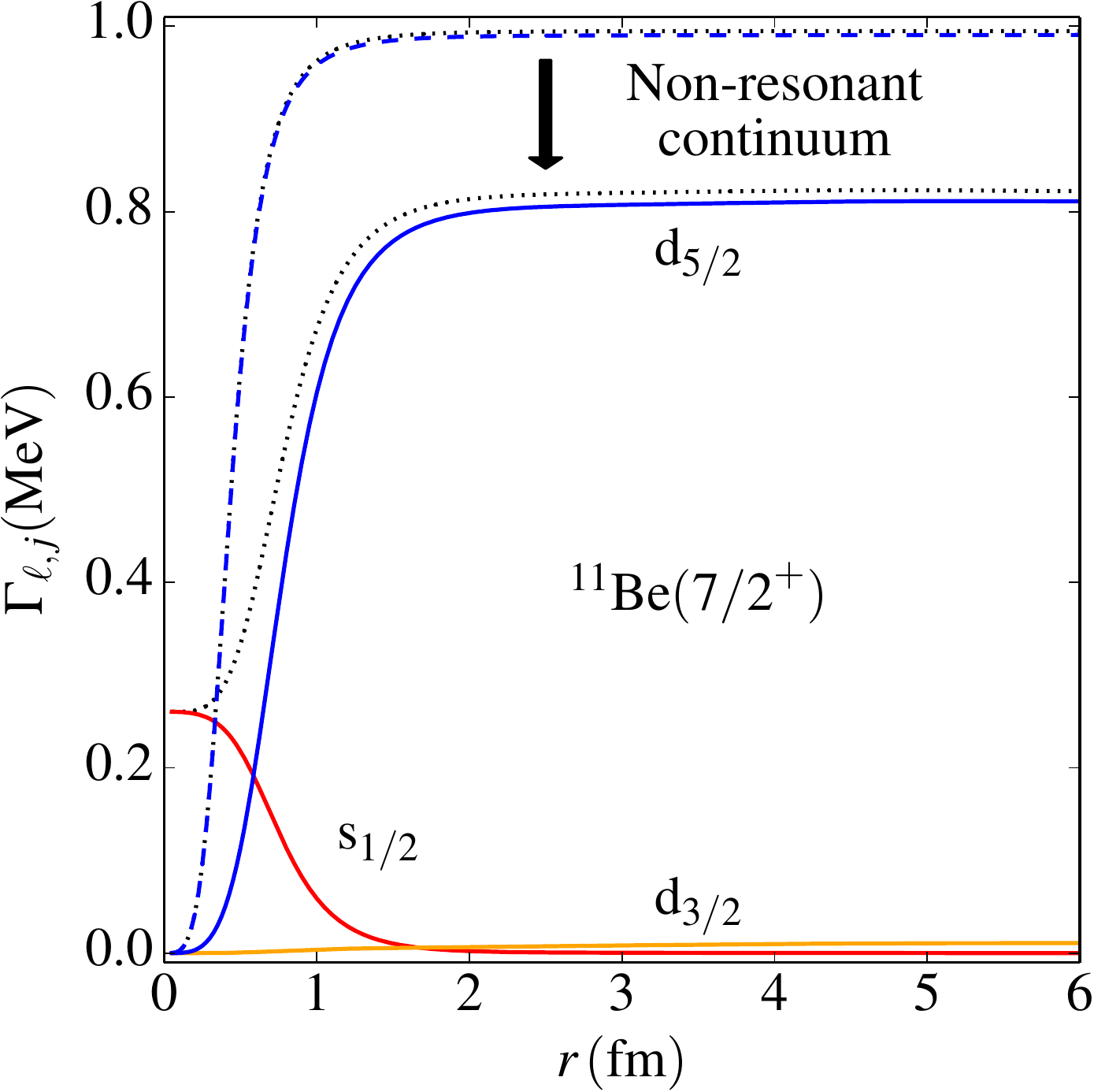}
		\caption{(Color online) Partial wave contribution to the width of the ${ {7/2}^{+} }$ yrast state of ${ {}^{11}\text{Be} }$. Continuous and dashed lines correspond to the full and pole space calculations, respectively. The total width is indicated with dotted line in both cases.}
		\label{fig_4}
	\end{figure}

The non-resonant continuum does impact the predicted widths of broader resonances. This is illustrated  in  Fig.~\ref{fig_4}, which displays the  partial wave contributions to the total width for the ${ {7/2}_1^+ }$ state. The calculations were performed for
 ${ {\ell}_{ \text{max} } = 6 }$ using the flux formula in BEM, in both the full space and the pole space, in which the non-resonant contour is ignored and the basis completeness is broken. It is seen that removing the non-resonant part from the Berggren ensemble yields a width that is about 17\% larger than the value obtained in the full basis.
Similar shifts can be observed for other states.

The above discussion does not hold for excited (yrare) states of $^{11}$Be with $J=1/2$ and $3/2$, for which the $\ell=0$ channel is not blocked.
As seen in Fig.~\ref{fig_3}(d), the total widths of those states is predicted to be around $0.4$\,MeV, and it is even totally dominated by  $s_{1/2}$ for $J=1/2$.
For the $J=5/2$ yrare state, the $\ell=0$ channel is blocked again; hence, its width is  small.

\textit{Conclusion} -- We have investigated the existence of rotational bands embedded in the particle continuum using a non-adiabatic  particle-plus-core model.
As case in point, we considered the yrast band of the one-neutron halo  nucleus $^{11}$Be built upon the $J^\pi=1/2^+$ ground state.
We show that strong Coriolis  effects result in a rotational alignment of the valence neutron.
Owing to their large $\ell$-content, and the fact that low-$\ell$ channels are blocked,  the levels forming the yrast line of $^{11}$Be are expected to have small one-neutron widths: the neutron decay is predicted to proceed primarily via the unfavored sequence.
While our conclusions have been obtained by means of a schematic model, they nicely complement large-scale no-core shell model  calculations \cite{maris15_1161,caprio13_1158}, in which a decoupled $K^\pi=1/2^+$ band appears in $^{11}$Be viewed as an 11-body closed quantum system.
Our results suggest that the coupling to the one-neutron continuum will  play a minor role in the microscopic description of the yrast sequence.
This is not the case for the yrare states of $^{11}$Be, for which the $s$-wave decay is allowed.
Here, the neutron  widths can be large and continuum coupling effects -- appreciable. 
Unlike in the case of dipolar anions, where the valence electron density in unbound states can be extremely diffused due to the shallowness of the electron-molecule pseudopotential, the calculated valence neutron  states in ${^{11}}$Be are fairly well localized within the range of the neutron-core potential.

In summary, we demonstrated that collective states can exist at high excitation energy in weakly bound neutron drip-line nuclei such as $^{11}$Be.
The calculated decay widths of those states can be narrow, and the resulting half-lives can be  long  compared to the single-particle time scale.
This justifies the use of a geometric picture in such cases.
However, whether the notion of collective nuclear states applies to very broad resonances with $T_{1/2}< T_{\rm s.p.}$ is still under discussion.
This matter will be a subject of forthcoming investigations.


\begin{acknowledgments}
Useful discussions with Eduardo Garrido, Michael Thoennessen,  and Ian Thompson are gratefully acknowledged.
This material is
based upon work supported by the U.S.\ Department of Energy, Office of
Science, Office of Nuclear Physics under  award numbers 
DE-SC0013365 (Michigan State University) and
DE-FG02-10ER41700 (French-U.S. Theory Institute for Physics with Exotic Nuclei).
\end{acknowledgments}


\begin{thebibliography}{42}%
\makeatletter
\providecommand \@ifxundefined [1]{%
 \@ifx{#1\undefined}
}%
\providecommand \@ifnum [1]{%
 \ifnum #1\expandafter \@firstoftwo
 \else \expandafter \@secondoftwo
 \fi
}%
\providecommand \@ifx [1]{%
 \ifx #1\expandafter \@firstoftwo
 \else \expandafter \@secondoftwo
 \fi
}%
\providecommand \natexlab [1]{#1}%
\providecommand \enquote  [1]{``#1''}%
\providecommand \bibnamefont  [1]{#1}%
\providecommand \bibfnamefont [1]{#1}%
\providecommand \citenamefont [1]{#1}%
\providecommand \href@noop [0]{\@secondoftwo}%
\providecommand \href [0]{\begingroup \@sanitize@url \@href}%
\providecommand \@href[1]{\@@startlink{#1}\@@href}%
\providecommand \@@href[1]{\endgroup#1\@@endlink}%
\providecommand \@sanitize@url [0]{\catcode `\\12\catcode `\$12\catcode
  `\&12\catcode `\#12\catcode `\^12\catcode `\_12\catcode `\%12\relax}%
\providecommand \@@startlink[1]{}%
\providecommand \@@endlink[0]{}%
\providecommand \url  [0]{\begingroup\@sanitize@url \@url }%
\providecommand \@url [1]{\endgroup\@href {#1}{\urlprefix }}%
\providecommand \urlprefix  [0]{URL }%
\providecommand \Eprint [0]{\href }%
\providecommand \doibase [0]{http://dx.doi.org/}%
\providecommand \selectlanguage [0]{\@gobble}%
\providecommand \bibinfo  [0]{\@secondoftwo}%
\providecommand \bibfield  [0]{\@secondoftwo}%
\providecommand \translation [1]{[#1]}%
\providecommand \BibitemOpen [0]{}%
\providecommand \bibitemStop [0]{}%
\providecommand \bibitemNoStop [0]{.\EOS\space}%
\providecommand \EOS [0]{\spacefactor3000\relax}%
\providecommand \BibitemShut  [1]{\csname bibitem#1\endcsname}%
\let\auto@bib@innerbib\@empty
\bibitem [{\citenamefont {Jensen}\ \emph {et~al.}(2004)\citenamefont {Jensen},
  \citenamefont {Riisager}, \citenamefont {Fedorov},\ and\ \citenamefont
  {Garrido}}]{jensen04_233}%
  \BibitemOpen
  \bibfield  {author} {\bibinfo {author} {\bibfnamefont {A.~S.}\ \bibnamefont
  {Jensen}}, \bibinfo {author} {\bibfnamefont {K.}~\bibnamefont {Riisager}},
  \bibinfo {author} {\bibfnamefont {D.~V.}\ \bibnamefont {Fedorov}}, \ and\
  \bibinfo {author} {\bibfnamefont {E.}~\bibnamefont {Garrido}},\ }\href
  {http://dx.doi.org/10.1103/RevModPhys.76.215} {\bibfield  {journal} {\bibinfo
   {journal} {Rev. Mod. Phys.}\ }\textbf {\bibinfo {volume} {76}},\ \bibinfo
  {pages} {215} (\bibinfo {year} {2004})}\BibitemShut {NoStop}%
\bibitem [{\citenamefont {Tanihata}\ \emph {et~al.}(2013)\citenamefont
  {Tanihata}, \citenamefont {Savajols},\ and\ \citenamefont
  {Kanungo}}]{tanihata13_549}%
  \BibitemOpen
  \bibfield  {author} {\bibinfo {author} {\bibfnamefont {I.}~\bibnamefont
  {Tanihata}}, \bibinfo {author} {\bibfnamefont {H.}~\bibnamefont {Savajols}},
  \ and\ \bibinfo {author} {\bibfnamefont {R.}~\bibnamefont {Kanungo}},\ }\href
  {http://dx.doi.org/10.1016/j.ppnp.2012.07.001} {\bibfield  {journal}
  {\bibinfo  {journal} {Prog. Part. Nucl. Phys.}\ }\textbf {\bibinfo {volume}
  {68}},\ \bibinfo {pages} {215} (\bibinfo {year} {2013})}\BibitemShut
  {NoStop}%
\bibitem [{\citenamefont {von Oertzen}\ \emph {et~al.}(2006)\citenamefont {von
  Oertzen}, \citenamefont {Freer},\ and\ \citenamefont
  {Kanada-En'yo}}]{oertzen06_1017}%
  \BibitemOpen
  \bibfield  {author} {\bibinfo {author} {\bibfnamefont {W.}~\bibnamefont {von
  Oertzen}}, \bibinfo {author} {\bibfnamefont {M.}~\bibnamefont {Freer}}, \
  and\ \bibinfo {author} {\bibfnamefont {Y.}~\bibnamefont {Kanada-En'yo}},\
  }\href {http://dx.doi.org/10.1016/j.physrep.2006.07.001} {\bibfield
  {journal} {\bibinfo  {journal} {Phys. Rep.}\ }\textbf {\bibinfo {volume}
  {432}},\ \bibinfo {pages} {43} (\bibinfo {year} {2006})}\BibitemShut
  {NoStop}%
\bibitem [{\citenamefont {Freer}(2007)}]{freer07_1018}%
  \BibitemOpen
  \bibfield  {author} {\bibinfo {author} {\bibfnamefont {M.}~\bibnamefont
  {Freer}},\ }\href {http://dx.doi.org/10.1088/0034-4885/70/12/R03} {\bibfield
  {journal} {\bibinfo  {journal} {Rep. Prog. Phys.}\ }\textbf {\bibinfo
  {volume} {70}},\ \bibinfo {pages} {2149} (\bibinfo {year}
  {2007})}\BibitemShut {NoStop}%
\bibitem [{\citenamefont {Oko{\l}owicz}\ \emph {et~al.}(2013)\citenamefont
  {Oko{\l}owicz}, \citenamefont {Nazarewicz},\ and\ \citenamefont
  {P{\l}oszajczak}}]{okolowicz13_241}%
  \BibitemOpen
  \bibfield  {author} {\bibinfo {author} {\bibfnamefont {J.}~\bibnamefont
  {Oko{\l}owicz}}, \bibinfo {author} {\bibfnamefont {W.}~\bibnamefont
  {Nazarewicz}}, \ and\ \bibinfo {author} {\bibfnamefont {M.}~\bibnamefont
  {P{\l}oszajczak}},\ }\href {http://dx.doi.org/10.1002/prop.201200127}
  {\bibfield  {journal} {\bibinfo  {journal} {Fortschr. Phys.}\ }\textbf
  {\bibinfo {volume} {61}},\ \bibinfo {pages} {66} (\bibinfo {year}
  {2013})}\BibitemShut {NoStop}%
\bibitem [{\citenamefont {Oko{\l}owicz}\ \emph {et~al.}(2012)\citenamefont
  {Oko{\l}owicz}, \citenamefont {P{\l}oszajczak},\ and\ \citenamefont
  {Nazarewicz}}]{okolowicz12_998}%
  \BibitemOpen
  \bibfield  {author} {\bibinfo {author} {\bibfnamefont {J.}~\bibnamefont
  {Oko{\l}owicz}}, \bibinfo {author} {\bibfnamefont {M.}~\bibnamefont
  {P{\l}oszajczak}}, \ and\ \bibinfo {author} {\bibfnamefont {W.}~\bibnamefont
  {Nazarewicz}},\ }\href {http://dx.doi.org/10.1143/PTPS.196.230} {\bibfield
  {journal} {\bibinfo  {journal} {Prog. Theor. Phys. Supp.}\ }\textbf {\bibinfo
  {volume} {196}},\ \bibinfo {pages} {230} (\bibinfo {year}
  {2012})}\BibitemShut {NoStop}%
\bibitem [{\citenamefont {Matsuo}\ and\ \citenamefont
  {Nakatsukasa}(2010)}]{Matsuo10}%
  \BibitemOpen
  \bibfield  {author} {\bibinfo {author} {\bibfnamefont {M.}~\bibnamefont
  {Matsuo}}\ and\ \bibinfo {author} {\bibfnamefont {T.}~\bibnamefont
  {Nakatsukasa}},\ }\href {http://stacks.iop.org/0954-3899/37/i=6/a=064017}
  {\bibfield  {journal} {\bibinfo  {journal} {J. Phys. G}\ }\textbf {\bibinfo
  {volume} {37}},\ \bibinfo {pages} {064017} (\bibinfo {year}
  {2010})}\BibitemShut {NoStop}%
\bibitem [{\citenamefont {Michel}\ \emph {et~al.}(2010)\citenamefont {Michel},
  \citenamefont {Nazarewicz}, \citenamefont {Oko{\l}owicz},\ and\ \citenamefont
  {P{\l}oszajczak}}]{michel10_4}%
  \BibitemOpen
  \bibfield  {author} {\bibinfo {author} {\bibfnamefont {N.}~\bibnamefont
  {Michel}}, \bibinfo {author} {\bibfnamefont {W.}~\bibnamefont {Nazarewicz}},
  \bibinfo {author} {\bibfnamefont {J.}~\bibnamefont {Oko{\l}owicz}}, \ and\
  \bibinfo {author} {\bibfnamefont {M.}~\bibnamefont {P{\l}oszajczak}},\ }\href
  {http://dx.doi.org/10.1088/0954-3899/37/6/064042} {\bibfield  {journal}
  {\bibinfo  {journal} {J. Phys. G: Nucl. Part. Phys.}\ }\textbf {\bibinfo
  {volume} {37}},\ \bibinfo {pages} {064042} (\bibinfo {year}
  {2010})}\BibitemShut {NoStop}%
\bibitem [{\citenamefont {Michel}\ \emph {et~al.}(2009)\citenamefont {Michel},
  \citenamefont {Nazarewicz}, \citenamefont {P{\l}oszajczak},\ and\
  \citenamefont {Vertse}}]{michel09_2}%
  \BibitemOpen
  \bibfield  {author} {\bibinfo {author} {\bibfnamefont {N.}~\bibnamefont
  {Michel}}, \bibinfo {author} {\bibfnamefont {W.}~\bibnamefont {Nazarewicz}},
  \bibinfo {author} {\bibfnamefont {M.}~\bibnamefont {P{\l}oszajczak}}, \ and\
  \bibinfo {author} {\bibfnamefont {T.}~\bibnamefont {Vertse}},\ }\href
  {http://dx.doi.org/10.1088/0954-3899/36/1/013101} {\bibfield  {journal}
  {\bibinfo  {journal} {J. Phys. G}\ }\textbf {\bibinfo {volume} {36}},\
  \bibinfo {pages} {013101} (\bibinfo {year} {2009})}\BibitemShut {NoStop}%
\bibitem [{\citenamefont {Born}\ and\ \citenamefont
  {Oppenheimer}(1927)}]{born27}%
  \BibitemOpen
  \bibfield  {author} {\bibinfo {author} {\bibfnamefont {M.}~\bibnamefont
  {Born}}\ and\ \bibinfo {author} {\bibfnamefont {R.}~\bibnamefont
  {Oppenheimer}},\ }\href {http://dx.doi.org/10.1002/andp.19273892002}
  {\bibfield  {journal} {\bibinfo  {journal} {Ann. Physik (Leipzig)}\ }\textbf
  {\bibinfo {volume} {389}},\ \bibinfo {pages} {457} (\bibinfo {year}
  {1927})}\BibitemShut {NoStop}%
\bibitem [{\citenamefont {Nazarewicz}(1993)}]{naz93}%
  \BibitemOpen
  \bibfield  {author} {\bibinfo {author} {\bibfnamefont {W.}~\bibnamefont
  {Nazarewicz}},\ }\href {\doibase
  http://dx.doi.org/10.1016/0375-9474(93)90565-F} {\bibfield  {journal}
  {\bibinfo  {journal} {Nucl. Phys. A}\ }\textbf {\bibinfo {volume} {557}},\
  \bibinfo {pages} {489 } (\bibinfo {year} {1993})}\BibitemShut {NoStop}%
\bibitem [{\citenamefont {Mar{\'i}n-L{\'a}mbarri}\ \emph
  {et~al.}(2014)\citenamefont {Mar{\'i}n-L{\'a}mbarri}, \citenamefont {Bijker},
  \citenamefont {Freer}, \citenamefont {Gai}, \citenamefont {Kokalova},
  \citenamefont {Parker},\ and\ \citenamefont {Wheldon}}]{Lambarri14}%
  \BibitemOpen
  \bibfield  {author} {\bibinfo {author} {\bibfnamefont {D.~J.}\ \bibnamefont
  {Mar{\'i}n-L{\'a}mbarri}}, \bibinfo {author} {\bibfnamefont {R.}~\bibnamefont
  {Bijker}}, \bibinfo {author} {\bibfnamefont {M.}~\bibnamefont {Freer}},
  \bibinfo {author} {\bibfnamefont {M.}~\bibnamefont {Gai}}, \bibinfo {author}
  {\bibfnamefont {T.}~\bibnamefont {Kokalova}}, \bibinfo {author}
  {\bibfnamefont {D.~J.}\ \bibnamefont {Parker}}, \ and\ \bibinfo {author}
  {\bibfnamefont {C.}~\bibnamefont {Wheldon}},\ }\href {\doibase
  10.1103/PhysRevLett.113.012502} {\bibfield  {journal} {\bibinfo  {journal}
  {Phys. Rev. Lett.}\ }\textbf {\bibinfo {volume} {113}},\ \bibinfo {pages}
  {012502} (\bibinfo {year} {2014})}\BibitemShut {NoStop}%
\bibitem [{\citenamefont {Goldanskii}(1966)}]{goldanskii66_1159}%
  \BibitemOpen
  \bibfield  {author} {\bibinfo {author} {\bibfnamefont {V.~I.}\ \bibnamefont
  {Goldanskii}},\ }\href
  {http://dx.doi.org/10.1146/annurev.ns.16.120166.000245} {\bibfield  {journal}
  {\bibinfo  {journal} {Ann. Rev. Nucl. Sci.}\ }\textbf {\bibinfo {volume}
  {16}},\ \bibinfo {pages} {1} (\bibinfo {year} {1966})}\BibitemShut {NoStop}%
\bibitem [{\citenamefont {Thoennessen}(2004)}]{thoennessen04_1165}%
  \BibitemOpen
  \bibfield  {author} {\bibinfo {author} {\bibfnamefont {M.}~\bibnamefont
  {Thoennessen}},\ }\href {http://dx.doi.org/10.1088/0034-4885/67/7/R04}
  {\bibfield  {journal} {\bibinfo  {journal} {Rep. Prog. Phys.}\ }\textbf
  {\bibinfo {volume} {67}},\ \bibinfo {pages} {1187} (\bibinfo {year}
  {2004})}\BibitemShut {NoStop}%
\bibitem [{\citenamefont {Garrido}\ \emph {et~al.}(2013)\citenamefont
  {Garrido}, \citenamefont {Jensen},\ and\ \citenamefont
  {Fedorov}}]{garrido13_1171}%
  \BibitemOpen
  \bibfield  {author} {\bibinfo {author} {\bibfnamefont {E.}~\bibnamefont
  {Garrido}}, \bibinfo {author} {\bibfnamefont {A.~S.}\ \bibnamefont {Jensen}},
  \ and\ \bibinfo {author} {\bibfnamefont {D.~V.}\ \bibnamefont {Fedorov}},\
  }\href {http://dx.doi.org/10.1103/PhysRevC.88.024001} {\bibfield  {journal}
  {\bibinfo  {journal} {Phys. Rev. C}\ }\textbf {\bibinfo {volume} {88}},\
  \bibinfo {pages} {024001} (\bibinfo {year} {2013})}\BibitemShut {NoStop}%
\bibitem [{\citenamefont {Fynbo}\ \emph {et~al.}(2009)\citenamefont {Fynbo},
  \citenamefont {{\'A}lvarez-Rodr{\'{i}}guez}, \citenamefont {Jensen},
  \citenamefont {Kirsebom}, \citenamefont {Fedorov},\ and\ \citenamefont
  {Garrido}}]{Fynbo09}%
  \BibitemOpen
  \bibfield  {author} {\bibinfo {author} {\bibfnamefont {H.~O.~U.}\
  \bibnamefont {Fynbo}}, \bibinfo {author} {\bibfnamefont {R.}~\bibnamefont
  {{\'A}lvarez-Rodr{\'{i}}guez}}, \bibinfo {author} {\bibfnamefont {A.~S.}\
  \bibnamefont {Jensen}}, \bibinfo {author} {\bibfnamefont {O.~S.}\
  \bibnamefont {Kirsebom}}, \bibinfo {author} {\bibfnamefont {D.~V.}\
  \bibnamefont {Fedorov}}, \ and\ \bibinfo {author} {\bibfnamefont
  {E.}~\bibnamefont {Garrido}},\ }\href {\doibase 10.1103/PhysRevC.79.054009}
  {\bibfield  {journal} {\bibinfo  {journal} {Phys. Rev. C}\ }\textbf {\bibinfo
  {volume} {79}},\ \bibinfo {pages} {054009} (\bibinfo {year}
  {2009})}\BibitemShut {NoStop}%
\bibitem [{\citenamefont {Riisager}\ \emph {et~al.}(2015)\citenamefont
  {Riisager}, \citenamefont {Fynbo}, \citenamefont {Hyldegaard},\ and\
  \citenamefont {Jensen}}]{Rii15}%
  \BibitemOpen
  \bibfield  {author} {\bibinfo {author} {\bibfnamefont {K.}~\bibnamefont
  {Riisager}}, \bibinfo {author} {\bibfnamefont {H.}~\bibnamefont {Fynbo}},
  \bibinfo {author} {\bibfnamefont {S.}~\bibnamefont {Hyldegaard}}, \ and\
  \bibinfo {author} {\bibfnamefont {A.}~\bibnamefont {Jensen}},\ }\href
  {\doibase http://dx.doi.org/10.1016/j.nuclphysa.2015.04.003} {\bibfield
  {journal} {\bibinfo  {journal} {Nucl. Phys. A}\ }\textbf {\bibinfo {volume}
  {940}},\ \bibinfo {pages} {119} (\bibinfo {year} {2015})}\BibitemShut
  {NoStop}%
\bibitem [{\citenamefont {Pastore}\ \emph {et~al.}(2014)\citenamefont
  {Pastore}, \citenamefont {Wiringa}, \citenamefont {Pieper},\ and\
  \citenamefont {Schiavilla}}]{pastore14_1183}%
  \BibitemOpen
  \bibfield  {author} {\bibinfo {author} {\bibfnamefont {S.}~\bibnamefont
  {Pastore}}, \bibinfo {author} {\bibfnamefont {R.~B.}\ \bibnamefont
  {Wiringa}}, \bibinfo {author} {\bibfnamefont {S.~C.}\ \bibnamefont {Pieper}},
  \ and\ \bibinfo {author} {\bibfnamefont {R.}~\bibnamefont {Schiavilla}},\
  }\href {http://dx.doi.org/10.1103/PhysRevC.90.024321} {\bibfield  {journal}
  {\bibinfo  {journal} {Phys. Rev. C}\ }\textbf {\bibinfo {volume} {90}},\
  \bibinfo {pages} {024321} (\bibinfo {year} {2014})}\BibitemShut {NoStop}%
\bibitem [{\citenamefont {Caprio}\ \emph {et~al.}(2013)\citenamefont {Caprio},
  \citenamefont {Maris},\ and\ \citenamefont {Vary}}]{caprio13_1158}%
  \BibitemOpen
  \bibfield  {author} {\bibinfo {author} {\bibfnamefont {M.~A.}\ \bibnamefont
  {Caprio}}, \bibinfo {author} {\bibfnamefont {P.}~\bibnamefont {Maris}}, \
  and\ \bibinfo {author} {\bibfnamefont {J.~P.}\ \bibnamefont {Vary}},\ }\href
  {http://dx.doi.org/10.1016/j.physletb.2012.12.064} {\bibfield  {journal}
  {\bibinfo  {journal} {Phys. Lett. B}\ }\textbf {\bibinfo {volume} {719}},\
  \bibinfo {pages} {179} (\bibinfo {year} {2013})}\BibitemShut {NoStop}%
\bibitem [{\citenamefont {Hagen}\ \emph {et~al.}(2006)\citenamefont {Hagen},
  \citenamefont {Hjorth-Jensen},\ and\ \citenamefont {Michel}}]{hagen06_14}%
  \BibitemOpen
  \bibfield  {author} {\bibinfo {author} {\bibfnamefont {G.}~\bibnamefont
  {Hagen}}, \bibinfo {author} {\bibfnamefont {M.}~\bibnamefont
  {Hjorth-Jensen}}, \ and\ \bibinfo {author} {\bibfnamefont {N.}~\bibnamefont
  {Michel}},\ }\href {http://dx.doi.org/10.1103/PhysRevC.73.064307} {\bibfield
  {journal} {\bibinfo  {journal} {Phys. Rev. C}\ }\textbf {\bibinfo {volume}
  {73}},\ \bibinfo {pages} {064307} (\bibinfo {year} {2006})}\BibitemShut
  {NoStop}%
\bibitem [{\citenamefont {Quaglioni}\ \emph {et~al.}(2012)\citenamefont
  {Quaglioni}, \citenamefont {Navr\'atil}, \citenamefont {Roth},\ and\
  \citenamefont {Horiuchi}}]{quaglioni12_738}%
  \BibitemOpen
  \bibfield  {author} {\bibinfo {author} {\bibfnamefont {S.}~\bibnamefont
  {Quaglioni}}, \bibinfo {author} {\bibfnamefont {P.}~\bibnamefont
  {Navr\'atil}}, \bibinfo {author} {\bibfnamefont {R.}~\bibnamefont {Roth}}, \
  and\ \bibinfo {author} {\bibfnamefont {W.}~\bibnamefont {Horiuchi}},\ }\href
  {http://dx.doi.org/10.1088/1742-6596/402/1/012037} {\bibfield  {journal}
  {\bibinfo  {journal} {J. Phys.: Conf. Ser.}\ }\textbf {\bibinfo {volume}
  {402}},\ \bibinfo {pages} {012037} (\bibinfo {year} {2012})}\BibitemShut
  {NoStop}%
\bibitem [{\citenamefont {Papadimitriou}\ \emph {et~al.}(2013)\citenamefont
  {Papadimitriou}, \citenamefont {Rotureau}, \citenamefont {Michel},
  \citenamefont {P{\l}oszajczak},\ and\ \citenamefont
  {Barrett}}]{papadimitriou13_441}%
  \BibitemOpen
  \bibfield  {author} {\bibinfo {author} {\bibfnamefont {G.}~\bibnamefont
  {Papadimitriou}}, \bibinfo {author} {\bibfnamefont {J.}~\bibnamefont
  {Rotureau}}, \bibinfo {author} {\bibfnamefont {N.}~\bibnamefont {Michel}},
  \bibinfo {author} {\bibfnamefont {M.}~\bibnamefont {P{\l}oszajczak}}, \ and\
  \bibinfo {author} {\bibfnamefont {B.~R.}\ \bibnamefont {Barrett}},\ }\href
  {http://dx.doi.org/10.1103/PhysRevC.88.044318} {\bibfield  {journal}
  {\bibinfo  {journal} {Phys. Rev. C}\ }\textbf {\bibinfo {volume} {88}},\
  \bibinfo {pages} {044318} (\bibinfo {year} {2013})}\BibitemShut {NoStop}%
\bibitem [{\citenamefont {Berggren}(1968)}]{berggren68_32}%
  \BibitemOpen
  \bibfield  {author} {\bibinfo {author} {\bibfnamefont {T.}~\bibnamefont
  {Berggren}},\ }\href {http://dx.doi.org/10.1016/0375-9474(68)90593-9}
  {\bibfield  {journal} {\bibinfo  {journal} {Nucl. Phys. A}\ }\textbf
  {\bibinfo {volume} {109}},\ \bibinfo {pages} {265} (\bibinfo {year}
  {1968})}\BibitemShut {NoStop}%
\bibitem [{\citenamefont {Fossez}\ \emph {et~al.}(2015)\citenamefont {Fossez},
  \citenamefont {Michel}, \citenamefont {Nazarewicz}, \citenamefont
  {P{\l}oszajczak},\ and\ \citenamefont {Jaganathen}}]{fossez15_1028}%
  \BibitemOpen
  \bibfield  {author} {\bibinfo {author} {\bibfnamefont {K.}~\bibnamefont
  {Fossez}}, \bibinfo {author} {\bibfnamefont {N.}~\bibnamefont {Michel}},
  \bibinfo {author} {\bibfnamefont {W.}~\bibnamefont {Nazarewicz}}, \bibinfo
  {author} {\bibfnamefont {M.}~\bibnamefont {P{\l}oszajczak}}, \ and\ \bibinfo
  {author} {\bibfnamefont {Y.}~\bibnamefont {Jaganathen}},\ }\href
  {http://dx.doi.org/10.1103/PhysRevA.91.012503} {\bibfield  {journal}
  {\bibinfo  {journal} {Phys. Rev. A}\ }\textbf {\bibinfo {volume} {91}},\
  \bibinfo {pages} {012503} (\bibinfo {year} {2015})}\BibitemShut {NoStop}%
\bibitem [{\citenamefont {Kruppa}\ \emph {et~al.}(2000)\citenamefont {Kruppa},
  \citenamefont {Barmore}, \citenamefont {Nazarewicz},\ and\ \citenamefont
  {Vertse}}]{kruppa00_581}%
  \BibitemOpen
  \bibfield  {author} {\bibinfo {author} {\bibfnamefont {A.~T.}\ \bibnamefont
  {Kruppa}}, \bibinfo {author} {\bibfnamefont {B.}~\bibnamefont {Barmore}},
  \bibinfo {author} {\bibfnamefont {W.}~\bibnamefont {Nazarewicz}}, \ and\
  \bibinfo {author} {\bibfnamefont {T.}~\bibnamefont {Vertse}},\ }\href
  {http://dx.doi.org/10.1103/PhysRevLett.84.4549} {\bibfield  {journal}
  {\bibinfo  {journal} {Phys. Rev. Lett.}\ }\textbf {\bibinfo {volume} {84}},\
  \bibinfo {pages} {4549} (\bibinfo {year} {2000})}\BibitemShut {NoStop}%
\bibitem [{\citenamefont {Barmore}\ \emph {et~al.}(2000)\citenamefont
  {Barmore}, \citenamefont {Kruppa}, \citenamefont {Nazarewicz},\ and\
  \citenamefont {Vertse}}]{barmore00_582}%
  \BibitemOpen
  \bibfield  {author} {\bibinfo {author} {\bibfnamefont {B.}~\bibnamefont
  {Barmore}}, \bibinfo {author} {\bibfnamefont {A.~T.}\ \bibnamefont {Kruppa}},
  \bibinfo {author} {\bibfnamefont {W.}~\bibnamefont {Nazarewicz}}, \ and\
  \bibinfo {author} {\bibfnamefont {T.}~\bibnamefont {Vertse}},\ }\href
  {http://dx.doi.org/10.1103/PhysRevC.62.054315} {\bibfield  {journal}
  {\bibinfo  {journal} {Phys. Rev. C}\ }\textbf {\bibinfo {volume} {62}},\
  \bibinfo {pages} {054315} (\bibinfo {year} {2000})}\BibitemShut {NoStop}%
\bibitem [{\citenamefont {Kruppa}\ and\ \citenamefont
  {Nazarewicz}(2004)}]{kruppa04_472}%
  \BibitemOpen
  \bibfield  {author} {\bibinfo {author} {\bibfnamefont {A.~T.}\ \bibnamefont
  {Kruppa}}\ and\ \bibinfo {author} {\bibfnamefont {W.}~\bibnamefont
  {Nazarewicz}},\ }\href {http://dx.doi.org/10.1103/PhysRevC.69.054311}
  {\bibfield  {journal} {\bibinfo  {journal} {Phys. Rev. C}\ }\textbf {\bibinfo
  {volume} {69}},\ \bibinfo {pages} {054311} (\bibinfo {year}
  {2004})}\BibitemShut {NoStop}%
\bibitem [{\citenamefont {Esbensen}\ and\ \citenamefont
  {Davids}(2000)}]{esbensen00_386}%
  \BibitemOpen
  \bibfield  {author} {\bibinfo {author} {\bibfnamefont {H.}~\bibnamefont
  {Esbensen}}\ and\ \bibinfo {author} {\bibfnamefont {C.~N.}\ \bibnamefont
  {Davids}},\ }\href {http://dx.doi.org/10.1103/PhysRevC.63.014315} {\bibfield
  {journal} {\bibinfo  {journal} {Phys. Rev. C}\ }\textbf {\bibinfo {volume}
  {63}},\ \bibinfo {pages} {014315} (\bibinfo {year} {2000})}\BibitemShut
  {NoStop}%
\bibitem [{\citenamefont {Nunes}\ \emph {et~al.}(1996)\citenamefont {Nunes},
  \citenamefont {Thompson},\ and\ \citenamefont {Johnson}}]{nunes96_1127}%
  \BibitemOpen
  \bibfield  {author} {\bibinfo {author} {\bibfnamefont {F.~M.}\ \bibnamefont
  {Nunes}}, \bibinfo {author} {\bibfnamefont {I.~J.}\ \bibnamefont {Thompson}},
  \ and\ \bibinfo {author} {\bibfnamefont {R.~C.}\ \bibnamefont {Johnson}},\
  }\href {http://dx.doi.org/10.1016/0375-9474(95)00398-3} {\bibfield  {journal}
  {\bibinfo  {journal} {Nucl. Phys. A}\ }\textbf {\bibinfo {volume} {596}},\
  \bibinfo {pages} {171} (\bibinfo {year} {1996})}\BibitemShut {NoStop}%
\bibitem [{\citenamefont {Descouvemont}(1997)}]{Descouvemont1997261}%
  \BibitemOpen
  \bibfield  {author} {\bibinfo {author} {\bibfnamefont {P.}~\bibnamefont
  {Descouvemont}},\ }\href {\doibase
  http://dx.doi.org/10.1016/S0375-9474(97)00015-8} {\bibfield  {journal}
  {\bibinfo  {journal} {Nucl. Phys. A}\ }\textbf {\bibinfo {volume} {615}},\
  \bibinfo {pages} {261} (\bibinfo {year} {1997})}\BibitemShut {NoStop}%
\bibitem [{\citenamefont {Fossez}\ \emph {et~al.}(2013)\citenamefont {Fossez},
  \citenamefont {Michel}, \citenamefont {Nazarewicz},\ and\ \citenamefont
  {P{\l}oszajczak}}]{fossez13_552}%
  \BibitemOpen
  \bibfield  {author} {\bibinfo {author} {\bibfnamefont {K.}~\bibnamefont
  {Fossez}}, \bibinfo {author} {\bibfnamefont {N.}~\bibnamefont {Michel}},
  \bibinfo {author} {\bibfnamefont {W.}~\bibnamefont {Nazarewicz}}, \ and\
  \bibinfo {author} {\bibfnamefont {M.}~\bibnamefont {P{\l}oszajczak}},\ }\href
  {http://dx.doi.org/10.1103/PhysRevA.87.042515} {\bibfield  {journal}
  {\bibinfo  {journal} {Phys. Rev. A}\ }\textbf {\bibinfo {volume} {87}},\
  \bibinfo {pages} {042515} (\bibinfo {year} {2013})}\BibitemShut {NoStop}%
\bibitem [{\citenamefont {Dykhne}\ and\ \citenamefont
  {Chaplik}(1961)}]{dykhne61_1041}%
  \BibitemOpen
  \bibfield  {author} {\bibinfo {author} {\bibfnamefont {A.~M.}\ \bibnamefont
  {Dykhne}}\ and\ \bibinfo {author} {\bibfnamefont {A.~V.}\ \bibnamefont
  {Chaplik}},\ }\href {http://www.jetp.ac.ru/cgi-bin/dn/e_013_05_1002.pdf}
  {\bibfield  {journal} {\bibinfo  {journal} {Sov. Phys. JETP}\ }\textbf
  {\bibinfo {volume} {13}},\ \bibinfo {pages} {1002} (\bibinfo {year}
  {1961})}\BibitemShut {NoStop}%
\bibitem [{\citenamefont {Gyarmati}\ and\ \citenamefont
  {Vertse}(1971)}]{gyarmati71_38}%
  \BibitemOpen
  \bibfield  {author} {\bibinfo {author} {\bibfnamefont {B.}~\bibnamefont
  {Gyarmati}}\ and\ \bibinfo {author} {\bibfnamefont {T.}~\bibnamefont
  {Vertse}},\ }\href {http://dx.doi.org/10.1016/0375-9474(71)90095-9}
  {\bibfield  {journal} {\bibinfo  {journal} {Nucl. Phys. A}\ }\textbf
  {\bibinfo {volume} {160}},\ \bibinfo {pages} {523} (\bibinfo {year}
  {1971})}\BibitemShut {NoStop}%
\bibitem [{\citenamefont {Davids}\ and\ \citenamefont
  {Esbensen}(2004)}]{davids04_562}%
  \BibitemOpen
  \bibfield  {author} {\bibinfo {author} {\bibfnamefont {C.~N.}\ \bibnamefont
  {Davids}}\ and\ \bibinfo {author} {\bibfnamefont {H.}~\bibnamefont
  {Esbensen}},\ }\href {http://dx.doi.org/10.1103/PhysRevC.69.034314}
  {\bibfield  {journal} {\bibinfo  {journal} {Phys. Rev. C}\ }\textbf {\bibinfo
  {volume} {69}},\ \bibinfo {pages} {034314} (\bibinfo {year}
  {2004})}\BibitemShut {NoStop}%
\bibitem [{\citenamefont {Saito}(1969)}]{saito69}%
  \BibitemOpen
  \bibfield  {author} {\bibinfo {author} {\bibfnamefont {S.}~\bibnamefont
  {Saito}},\ }\href {http://dx.doi.org/10.1143/PTP.41.705} {\bibfield
  {journal} {\bibinfo  {journal} {Prog. Theor. Phys.}\ }\textbf {\bibinfo
  {volume} {41}},\ \bibinfo {pages} {705} (\bibinfo {year} {1969})}\BibitemShut
  {NoStop}%
\bibitem [{\citenamefont {{http://www.nndc.bnl.gov/ensdf}}(2015)}]{ensdf}%
  \BibitemOpen
  \bibfield  {author} {\bibinfo {author} {\bibnamefont
  {{http://www.nndc.bnl.gov/ensdf}}},\ }\href@noop {} {} (\bibinfo {year}
  {2015})\BibitemShut {NoStop}%
\bibitem [{\citenamefont {von Oertzen}(1998)}]{voer98}%
  \BibitemOpen
  \bibfield  {author} {\bibinfo {author} {\bibfnamefont {W.}~\bibnamefont {von
  Oertzen}},\ }\href
  {http://www.actaphys.uj.edu.pl/_old/vol29/abs/v29p0247.htm} {\bibfield
  {journal} {\bibinfo  {journal} {Acta Phys. Pol. B}\ }\textbf {\bibinfo
  {volume} {29}},\ \bibinfo {pages} {247} (\bibinfo {year} {1998})}\BibitemShut
  {NoStop}%
\bibitem [{\citenamefont {Bohlen}\ \emph {et~al.}(2008)\citenamefont {Bohlen},
  \citenamefont {von Oertzen}, \citenamefont {Kalpakchieva}, \citenamefont
  {Massey}, \citenamefont {Dorsch}, \citenamefont {Milin}, \citenamefont
  {Schulz}, \citenamefont {Kokalova},\ and\ \citenamefont
  {Wheldon}}]{Bohlen08}%
  \BibitemOpen
  \bibfield  {author} {\bibinfo {author} {\bibfnamefont {H.~G.}\ \bibnamefont
  {Bohlen}}, \bibinfo {author} {\bibfnamefont {W.}~\bibnamefont {von Oertzen}},
  \bibinfo {author} {\bibfnamefont {R.}~\bibnamefont {Kalpakchieva}}, \bibinfo
  {author} {\bibfnamefont {T.~N.}\ \bibnamefont {Massey}}, \bibinfo {author}
  {\bibfnamefont {T.}~\bibnamefont {Dorsch}}, \bibinfo {author} {\bibfnamefont
  {M.}~\bibnamefont {Milin}}, \bibinfo {author} {\bibfnamefont
  {C.}~\bibnamefont {Schulz}}, \bibinfo {author} {\bibfnamefont
  {T.}~\bibnamefont {Kokalova}}, \ and\ \bibinfo {author} {\bibfnamefont
  {C.}~\bibnamefont {Wheldon}},\ }\href
  {http://stacks.iop.org/1742-6596/111/i=1/a=012021} {\bibfield  {journal}
  {\bibinfo  {journal} {J. Phys. Conf. Ser.}\ }\textbf {\bibinfo {volume}
  {111}},\ \bibinfo {pages} {012021} (\bibinfo {year} {2008})}\BibitemShut
  {NoStop}%
\bibitem [{\citenamefont {Descouvemont}(2002)}]{descouvemont02_1129}%
  \BibitemOpen
  \bibfield  {author} {\bibinfo {author} {\bibfnamefont {P.}~\bibnamefont
  {Descouvemont}},\ }\href {http://dx.doi.org/10.1016/S0375-9474(01)01286-6}
  {\bibfield  {journal} {\bibinfo  {journal} {Nucl. Phys. A}\ }\textbf
  {\bibinfo {volume} {699}},\ \bibinfo {pages} {463} (\bibinfo {year}
  {2002})}\BibitemShut {NoStop}%
\bibitem [{\citenamefont {Maris}\ \emph {et~al.}(2015)\citenamefont {Maris},
  \citenamefont {Caprio},\ and\ \citenamefont {Vary}}]{maris15_1161}%
  \BibitemOpen
  \bibfield  {author} {\bibinfo {author} {\bibfnamefont {P.}~\bibnamefont
  {Maris}}, \bibinfo {author} {\bibfnamefont {M.~A.}\ \bibnamefont {Caprio}}, \
  and\ \bibinfo {author} {\bibfnamefont {J.~P.}\ \bibnamefont {Vary}},\ }\href
  {http://dx.doi.org/10.1103/PhysRevC.91.014310} {\bibfield  {journal}
  {\bibinfo  {journal} {Phys. Rev. C}\ }\textbf {\bibinfo {volume} {91}},\
  \bibinfo {pages} {014310} (\bibinfo {year} {2015})}\BibitemShut {NoStop}%
\bibitem [{\citenamefont {Humblet}\ and\ \citenamefont
  {Rosenfeld}(1961)}]{humblet61_174}%
  \BibitemOpen
  \bibfield  {author} {\bibinfo {author} {\bibfnamefont {J.}~\bibnamefont
  {Humblet}}\ and\ \bibinfo {author} {\bibfnamefont {L.}~\bibnamefont
  {Rosenfeld}},\ }\href {http://dx.doi.org/10.1016/0029-5582(61)90207-3}
  {\bibfield  {journal} {\bibinfo  {journal} {Nucl. Phys.}\ }\textbf {\bibinfo
  {volume} {26}},\ \bibinfo {pages} {529} (\bibinfo {year} {1961})}\BibitemShut
  {NoStop}%
\bibitem [{\citenamefont {Nazarewicz}\ \emph {et~al.}(2001)\citenamefont
  {Nazarewicz}, \citenamefont {Dobaczewski}, \citenamefont {Matev},
  \citenamefont {Mizutori},\ and\ \citenamefont {Satula}}]{nazarewicz01_1180}%
  \BibitemOpen
  \bibfield  {author} {\bibinfo {author} {\bibfnamefont {W.}~\bibnamefont
  {Nazarewicz}}, \bibinfo {author} {\bibfnamefont {J.}~\bibnamefont
  {Dobaczewski}}, \bibinfo {author} {\bibfnamefont {M.}~\bibnamefont {Matev}},
  \bibinfo {author} {\bibfnamefont {S.}~\bibnamefont {Mizutori}}, \ and\
  \bibinfo {author} {\bibfnamefont {W.}~\bibnamefont {Satula}},\ }\href
  {http://www.actaphys.uj.edu.pl/_old/vol32/pdf/v32p2349.pdf} {\bibfield
  {journal} {\bibinfo  {journal} {Acta Phys. Pol. B}\ }\textbf {\bibinfo
  {volume} {32}},\ \bibinfo {pages} {2349} (\bibinfo {year}
  {2001})}\BibitemShut {NoStop}%
\end{thebibliography}

%

\end{document}